\documentclass[tighten,twocolumn]{aastex701}
\usepackage[]{graphicx}
\usepackage{amsmath}
\usepackage[normalem]{ulem}

\newcommand{\forb}{f_{\mathrm{orb}}}
\newcommand{\fprec}{f_{\mathrm{prec}}}
\newcommand{\dinv}{\,d$^{-1}$}
\newcommand{\dlogl}{\delta\log\mathcal{L}}

\shortauthors{Kerr}
\shorttitle{$\gamma$-ray pulsar variability}

\begin{document}

\title{$\gamma$-ray Pulsar Emission is Mostly Stable on
Timescales from Minutes to Years}

\author[0000-0002-0893-4073,gname=Matthew,sname=Kerr]{M.~Kerr}
\affiliation{Space Science Division, Naval Research Laboratory, Washington, DC 20375--5352, USA}
\email{matthew.kerr@gmail.com}
\correspondingauthor{M.~Kerr}

\begin{abstract}

We present a method for the detection and characterization of random
  changes in the flux from $\gamma$-ray pulsars on sub-hour
  timescales, much shorter than variations that can be accessed using
  direct flux measurements.  Flux variations are a proxy for the
  variations in spindown power ($\dot{E}$) or particle acceleration, which can be
  produced by random switches between quasi-stable configurations
  of the pulsar magnetosphere.  This technique therefore probes the
  stability of pulsar magnetospheres and discrete spindown states on
  timescales much shorter than can be achieved with pulsar timing.  We
  apply the method to a sample of 115 bright $\gamma$-ray pulsars,
  finding no new instances of state changes.  We derive the
  sensitivity of the method and find that, for a wide range of
  possible state changing models, over a wide range of timescales, we
  can limit the amplitude of flux ($\dot{E}$) variations to $<$10\%.
  Substantial nulling is excluded in nearly all cases.  The best cases
  limit variations of any sort to $\leq$1\%.  These results indicate
  that $\gamma$-ray pulsar magnetospheres maintain a single
  configuration or narrow range of configurations with nearly constant
  power output.
\end{abstract}

\section{Introduction}
\label{sec:intro}

The theory of the structure, dynamics, and radiation of pulsar
magnetospheres has advanced tremendously since the turn of the
millennium.  Some major developments are:
\begin{enumerate}
  \item the determination of the global structure of the currents and
    magnetic field of a force-free (FF) magnetosphere
    \citep[e.g.][]{Contopoulos99,Spitkovsky06}, which results when there is
    sufficient co-rotating plasma to neutralize induced electric
    fields \citep{goldreich_pulsar_1969};
  \item the realization \citep{Timokhin06,beloborodov_polar-cap_2008}
    that supporting the global currents of the FF magnetosphere
    requires episodic formation of $e^{-}$/$e^{+}$ pairs
    just above the polar cap, with short bursts of particle
    acceleration in a strong vacuum field that is subsequently shorted
    out by the neutral\footnote{As a bonus, the fluctuations in the
    electric field as newly-produced plasma flows to neutralize the
    large-scale vacuum field may excite the oscillations that produce
    radio emission \citep[e.g.][]{philippov_origin_2020}.} pair plasma \citep{Timokhin10,Timokhin13}; the global currents moreover uniquely determine the pair-forming field lines \citep{gralla_inclined_2017}; 
  \item the validation with global particle-in-cell (PIC) simulations that
    the injection of pairs under a range of physically-motivated
    prescriptions results in a self-consistent force-free
    magnetosphere \citep[e.g.][]{Chen14,philippov_ab_2014,Cerutti15,Brambilla18};
  \item the more recent exploration of instabilities and reconnection
    in the equatorial current sheet at and beyond the light cylinder, where a substantial fraction of the Poynting
    flux is dissipated \citep[e.g.][]{Philippov15a}; these dynamic
    processes may directly accelerate pair-
    and GeV-producing particles via relativistic magnetic reconnection
    and Fermi acceleration via interaction with plasmoids
    \citep{hakobyan_magnetic_2023,soudais_scaling_2024}.
\end{enumerate}

The theoretical renaissance---see \citet{philippov_pulsar_2022} for an
excellent review---was underpinned by
$\gamma$-ray data provided by \textit{Fermi} LAT \citep{Atwood09}.  Because
$\gamma$-rays directly track the particle acceleration and comprise 1--30\% of
the spindown luminosity \citep{Smith23}, they uniquely convey the structure of the magnetosphere and constrain the macro- and
microphysics of accelerating regions.  Reproducing the observed pulse
profiles and energetics of $\gamma$-ray pulsars is a sine qua non for a
global magnetosphere model.

Despite this inspiring progress,  there is at least
one elephant in the room: whence pulsar state changing?  In the context of
simulations, it has been shown that differing the location and degree
of pair injection can successfully produce stable electrosphere and
force-free magnetosphere configurations \citep{Cerutti15,Brambilla18},
while some injection rates support a magnetosphere that flickers
between these states \citep{Chen20} on timescales of the pulse period\footnote{More precisely, the transit time of Alfv\'{e}n waves from
the light cylinder to the star.}, $P$.  Fluctuations in the location of the
Y-point \citep{Spitkovsky06}, as well as ``waving'' of the current
sheet due to the kink instability, are expected on similarly short
time scales \citep[e.g.][]{Philippov18}.  In essence, once a pulsar
achieves a force-free magnetosphere, its bulk properties ought to be
fixed on timescales $\gg$P.

However, this is inconsistent with observation: pulsars exhibit a
broad array of state changes.  The most dramatic of these are the
intermittent pulsars \citep{Kramer06}, of which about a half dozen are
known \citep[e.g.][]{camilo_psr_2012,lyne_two_2017}.  In these, the
radio pulse vanishes for up to years at a time.
When the radio pulse is absent, $\dot{\nu}$ drops by 50--100\%,
indicating that the magnetosphere has likely transitioned to an
electrosphere \citep{Krause-Polstorff85,Spitkovsky02}.

Also on month-to-year timescales, \citet{Lyne10} found much more
modest 0.1--10\% variations in $\dot{\nu}$ associated with changes in
the radio profile.  The variability of the pulse profile strongly
links the changes in $\dot{\nu}$ to the magnetosphere rather than,
e.g., spin irregularities originating in the neutron star itself
\citep[e.g.][]{melatos_pulsar_2014}.  Such linked variations,
including nearly periodic $\dot{\nu}$ variations \citep{Kerr16},
also occur in high-$\dot{E}$ pulsars
\citep{shaw_long-term_2022,lower_ubiquity_2025}.  In all of these
cases, there appears to be a switching timescale associated with each
pulsar, i.e., the switches are quasi-periodic, although there are
also suggestions that it is less common and more variable than in
lower $\dot{E}$ pulsars
\citep{brook_emission-rotation_2016,lower_rotational_2023}.

On much shorter timescales (from a few $P$ to minutes to hours), mode
changing---switching between one or more radio pulse profiles---has
long been known \citep[e.g.][]{Wang07}.  Nulling is a special case of
mode changing in which one of the modes is undetectable.  For these
pulsars, the switching is too rapid to measure any potential change in
$\dot{\nu}$.  However, for at least PSR~B0943$+$10, correlated changes
in radio and thermal X-ray emission suggests a substantial
change in the volume of the closed magnetosphere
\citep{hermsen_synchronous_2013}, while no such X-ray variations
accompany the dramatic radio mode changes of PSR~B1822$-$09
\citep{hermsen_simultaneous_2017}.

To isolate a potential physical origin of this ubiquitous state
changing, it is important to know if these slow (months) and fast
(minutes) state changes have the same underlying cause.
Analysis of the statistical properties of the state changes
\citep{cordes_pulsar_2013,Kerr14} indicates many examples demonstrate a
near-Markov property, supporting the idea that magnetospheres do
indeed accommodate metastable, force-free states with
durations ranging from seconds to years; and that global transitions
of the magnetosphere occur with a spectrum ranging from modest
reconfiguration to electrosphere transition, even on short timescales.

Direct measurements of $\dot{\nu}$ variations on short timescales
could establish a common origin and reveal the extent of the
magnetospheric variations, but such measurements are not possible with
pulsar timing.  As the primary (visible) exhaust of the pulsar
machine, $\gamma$ rays could potentially provide such a probe.  This
approach is motivated by the one (and only) known state-changing
$\gamma$-ray pulsar, PSR~J2021$+$4026 \citep{Allafort13}.  It exhibits
step-like, correlated variations in $\dot{\nu}$ and $\gamma$-ray flux
of about 10\% every 2--3 years \citep{fiori_phase-resolved_2024}.  For
this pulsar, $\gamma$-ray luminosity is a good tracer\footnote{For the sake of a simple interpretation, it is a
pity that the $\gamma$-ray flux and $\dot{E}$ are anti-correlated.} for
$\dot{E}$.
These variations have been attributed to changes in the magnetic field
configuration, which may be supported by a large phase shift of the
X-ray pulse profile
\citep{razzano_multiwavelength_2023} between the two states.

While \citet{Kerr22} was able to demonstrate the stability of the
magnetosphere of the few brightest $\gamma$-ray pulsars on
single-pulse timescales, a much broader parameter space remains to be
probed.  Here, with the view of diagnosing magnetospheric variations, we
develop a new methodology that enables the detections of
J2021$+$4206-like variations on much shorter timescales, down to tens
of minutes, and apply it to a sample of the 115 brightest
$\gamma$-ray pulsars.  This work thus bridges existing slow
variability searches like performed in the process of building the
Fourth \textit{Fermi} LAT Catalog \citep[4FGL, ][]{4FGL} and the
single-pulse variability searches of \citet{Kerr22}.  We present the
approach and methodology in \S\ref{sec:model} and
\S\ref{sec:methodology} and the specifics of the application to
\textit{Fermi} LAT data (\S\ref{sec:data}).  Although our focus is on
short timescales, we give results for both slow (\S\ref{sec:slow}) and
fast (\S\ref{sec:fast}) variability, and we summarize and discuss
these results in \S\ref{sec:discussion}.

\section{Strategy and Fast Variability Model}
\label{sec:model}

For a typical LAT pulsar, it is possible to make a $\gamma$-ray flux
measurement using about one month of data, so variability on monthly
timescales and longer can be directly characterized by comparing light
curves to the null hypothesis in a model-agnostic fashion, and such
searches have been carried out on all (4FGL) cataloged LAT sources,
including most pulsars.

On shorter time scales, and especially for fainter sources, it is no
longer possible to make point flux measurements.  In contrast, Fourier
methods excel because there are many realizations of fast
(high-frequency) components in a long data set, allowing detection of
very faint, periodic variability.  If---guided by the apparent Markov
property of many examples \citep{cordes_pulsar_2013}---we assume that
the fast variability originates in a stationary process, then it can
be represented via a spectrum $P(f)$.  We can then search for this
variability by computing $P(f)$ from the data and then determining if
it is consistent with the null hypothesis of a constant Poisson rate.

Our strategy then is to search for variability using two stages.
First, we analyze the data on long timescales and filter out any slow
variabilility.  We then search for fast variability in the filtered
data using Fourier methods and Markov process models.  We detail this
two-stage process further in \S\ref{sec:methodology}.

For the faster variability, we adopt a two-state quasiperiodic model
that captures the bulk of the state-switching phenomena discussed in
\S\ref{sec:intro}.
Let the pulsar reside in a faint (bright) state with flux $F_f$
($F_b$) for an average time $T_f$ ($T_b$).  This is an asymmetric
square wave with frequency $f=1/(T_f+T_b)$.  Because we work in
relative flux, the parameters satisfy the constraint $(T_f\,F_f +
T_b\,F_b)/(T_f + T_b)=1$.
The modulation factor governs the strength:
\begin{equation}
  \label{eq:M}
  \mathcal{M}\equiv \frac{F_b-F_f}{F_b+F_f}.
\end{equation}
$\mathcal{M}\in[0,1]$ with $\mathcal{M}=0$ indicating no flux contrast
between the states, while $\mathcal{M}=1$ indicates total nulling.
The goal of the searches, then, is to 
 find $\mathcal{M}>0$ or to demonstrate stability by limiting
$\mathcal{M}$ to small values.  We define the process asymmetry
analagously as
\begin{equation}
  \label{eq:A}
  \mathcal{A}\equiv \frac{T_b-T_f}{T_b+T_f},
\end{equation}
with $\mathcal{A}\in(-1,1]$.  $\mathcal{A}=0$ corresponds to a
symmetric square wave. As $\mathcal{A}\rightarrow-1$, a source is almost
always in the faint state, with very brief, bright pulses that
maximize the flux variance.  In contrast, a source with 
$\mathcal{A}\rightarrow1$ is almost always in the
bright state with very occasional flux dropouts.  This
case has low relative variance.  More concretely, total fractional
variance of the process over many cycles is:
\begin{equation}
  \label{eq:total_power}
  \sigma^2=\frac{\mathcal{M}^2\left(1-\mathcal{A}^2\right)}{\left(1+\mathcal{A}\mathcal{M}\right)^2}.
\end{equation}
For total nulling, this expression simplifies to $\sigma^2=(1-\mathcal{A})/(1+\mathcal{A})$,
demonstrating the low (infinite) variance in the
$\mathcal{A}\rightarrow1$ ($\mathcal{A}\rightarrow-1$) limits.
This
distinction will be important when we connect limits on the total power
($\sigma$) to $\mathcal{M}$.

The shape of the power spectrum depends only on $|\mathcal{A}|$.  The $\mathcal{A}=0$ square wave has a power spectral density (PSD) with
most of its power in the fundamental frequency $f$, and a rapidly
diminishing amount at higher odd harmonics: $P(nf)\propto
(2n-1)^{-2}$.  Increasing asymmetry corresponds
to a larger fraction of time spent in the faint (or bright) state,
and relatively more power at higher harmonics.
% The variance is worked out as follows:
% <F^2> = (TbFb^2 + TfFf^2)/(Tb+Tf) = Fb^2 (1+S*T) / (1+T)^2
% where S=Ff/Fb, T=Tf/Tb are the ratios defined in the text
% F0 = (TbFb +TfFf)/(Tb+Tf) = Fb*(1+ST)/(1+T)
% <F^2> - F0^2 = Fb^2(1-S)^2*T/(1+T)^2
% (this followed some algebra)
% Finally, plug in Fb = F0*(1+T)/(1+ST) to get
% <(F-<F>)^2>/<F>^2 = (1-S)^2*T/(1+ST)^2

% Or, in terms of these variables
% C = (Fb-Ff)/(Fb+Ff)
% A = (Tb-Tf)/(Tb+Tf)
% <F^2> = (TbFb^2 + TfFf^2)/(Tb+Tf) = 1/2 [(1+A)Fb^2 + (1-A)Ff^2]
% <F0>^2 = [(TbFb +TfFf)/(Tb+Tf)]^2 = 1/4 [(1+A)Fb + (1-A)Ff]^2
% <F^2> - <F0>^2 = 1/4 (1-A^2)(Fb-Ff)^2
% <F^2>/F0^2 - 1 = (1-A^2)(Fb-Ff)^2 / [(1+A)Fb + (1-A)Ff]^2
% = C^2(1-A^2) / (1+AC)^2

We complete the model with a random element: the durations for each
faint and bright state are drawn from a uniform distribution of
half-width $W_f$ and $W_b$, respectively.  These parameters obey the
constraints $0 \leq W_f \leq T_f$ and $0 \leq W_b \leq T_b$.  $W_f=W_b=0$ is the deterministic,
asymmetric square wave.  In the maximally random case, $W_f=T_f$, so
that the faint state durations range from $0$ to $2T_f$, with the
inclusion of 0 indicating the model can accommodate ``skipping'' a
state.  The random elements of the model degrade the periodicity,
blurring the sharp features in $P(f)$ by an amount related to
\begin{equation}
  Q\equiv\frac{T_f+T_b}{W_f+W_b},
\end{equation}which is analogous to a cavity
quality factor.  Large $Q$ yields a $P(f)$ dominated by sharp lines
while $Q=1$ is maximally random, yielding a PSD with broadband
noise, as we will demonstrate below.

\begin{figure}
\centering
  \includegraphics[angle=0,width=0.98\linewidth]{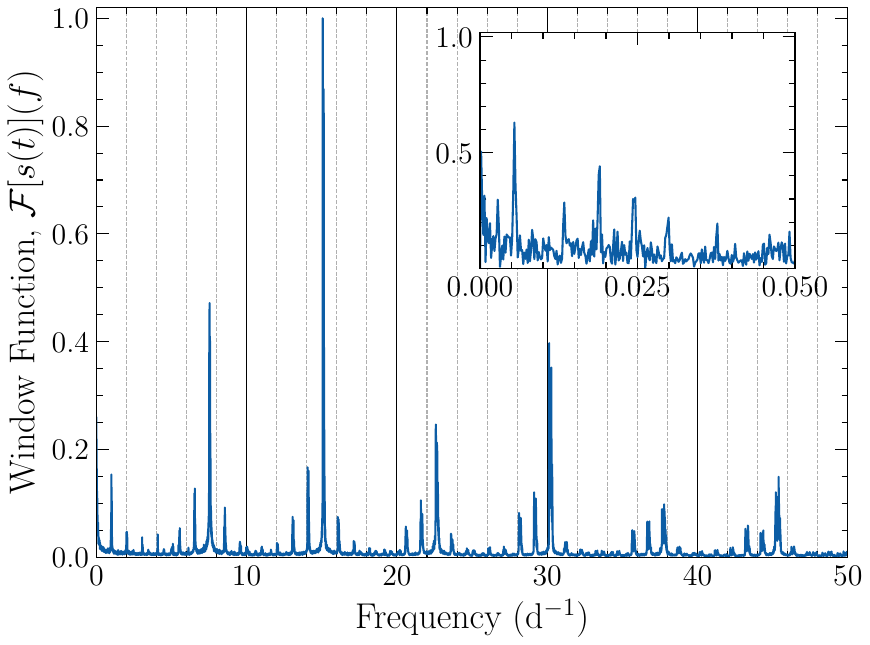}
  \caption{\label{fig:window}The window function as realized by the
  Fourier transform of the expected source counts, $s(t)$,
  towards PSR~J0633$+$1746.  The dominant peaks occur at the
  spacecraft orbital frequency, $f\approx15.1$\,\dinv, at
  $f\approx1$\,\dinv, and at beats of these frequencies.  The inset
  shows low-frequency power, including at the spacecraft precessional
  period, $\fprec \approx 0.019$\dinv{}.  We
  emphasize that there is no intrinsic source variability:  the
  peaks in the window function are due entirely to exposure
  modulation, and can be viewed as spectral leakage from $f=0$ to
  $f>0$.}
\end{figure}

\begin{figure*}
\centering
  \includegraphics[angle=0,width=0.98\linewidth]{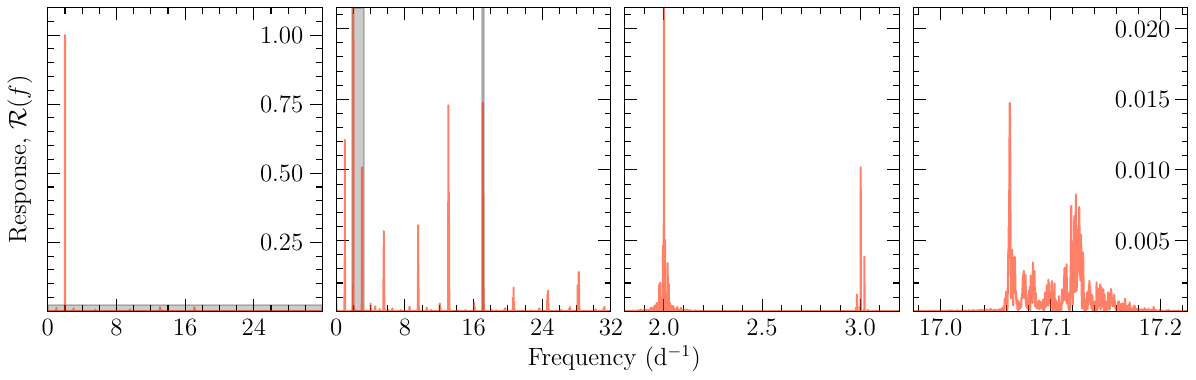}
  \caption{\label{fig:harm_example}The frequency response $R(f)$ to a
  $f=2$\dinv{} pure sinusoid for 15.6\,yr of data towards PSR~J0633$+$1746, normalized to a peak value of 1.  The first panel demonstrates the
  dominance of the main signal lobe.  The grey region with
  $R(f)<0.0215$ is shaded and expanded in the second panel to show the
  1--2\% spectral leakage to a wide range of frequencies.  The grey
  regions in this panel are further expanded in panels three and four.  The three rightmost panels all share a common y-axis range.
  The third panel shows leakage around the main lobe and a triplet
  formed at $f+1$\dinv{} and further split by $\fprec \approx
  0.019$\dinv{}, the spacecraft precession frequency.  The fourth panel shows leakage to
  $\forb\approx15.1$\dinv{}, which forms a broad forest due to
  evolution of $\forb$ over time.  Leakage to $f-\forb$ wraps
  around the origin to $f=$13.1\dinv{}, visible in the second panel.}
\end{figure*}

\section{Methodology}
\label{sec:methodology}

The implementation of the two-stage variability search strategy relies heavily on
techniques developed in \citet[][hereafter K19]{Kerr19}, and we
briefly review that connection.

Because of the degree-scale angular resolution of the LAT
\citep{Ajello21_onorbit}, sources are blended with diffuse emission
and other nearby point sources, making it impossible to select a
sample of photons ``from'' a given source.  Instead, for applications
such as pulsation searches, it is best to select a large, PSF-sized,
background-contaminated sample and assign to each photon a weight.
\citet{Bickel08} showed that these weights should be the probability
that a photon originates from the source of interest, and
\citet{Kerr11} demonstrated that the construction and application of
such weights can boost the sensitivity of pulsation/variability
searches.  These probabilities can be obtained with a model of sources
in the $\gamma$-ray sky, such as from 4FGL along with the instrument
response function, allowing the computation of the expected
flux for each source as a function of time, energy, and position.  The
weight is the ratio of the source flux to the total background flux
for the reconstructed time, energy, and position of each photon.

Consequently, the weights contain a distilled version of the sky
model.  While they are computed assuming one realization of the source
intensity, $\alpha_0(t)$ (typically constant for pulsars), the weights
can be exactly adjusted post facto for some different variability
pattern, $\alpha(t)$.  Conversely, the observed distribution of weights over
time can be used to make inferences about flux variations, and
K19 derived several computationally efficient methods to
estimate $\alpha(t)$ relative to $\alpha_0(t)$.  The same approach can
be taken to make inferences about variations of the total background,
$\beta(t)$.\footnote{Because it is a superposition of variable and
steady sources, the total background does not vary coherently.
However, it is often the case that a single background source dominates the total
variability, and that its spectrum is similar to other background
sources.  In this case, jointly estimating $\alpha(t)$ and
$\beta(t)$ allows separation of source and background variability.}

In more detail, the source rate is evaluated from the instantaneous instrument exposure and a model for the
source spectrum and integrated over cells of uniform length---say
300\,s---to obtain a time series for the predicted counts in each
cell, $s(t,E)$.  This quantity can be compared to the binned weights,
$w(t,E)$.  We define the total source counts $S=\int\,dE\,dt s(t,E)$.  The
background time series is simply $1-w(t,E)$, and the background counts
are obtained using the total source-to-background ratio as $b(t,E) =
s(t,E)\times\frac{S}{N-S}$.  These time series are the ingredients to
the likelihood functions for estimating $\alpha(t)$ and $\beta(t)$

\subsection{Slow Variability With Bayesian Blocks}
\label{sec:matched}

Two methods are particularly useful here.  First,
K19 demonstrated an application of the Bayesian blocks (BB)
algorithm \citep{Scargle98,Scargle13} directly to the photon weights.
The BB algorithm aggregates individual cells into longer
blocks that maximize a penalized likelihood function, yielding
optimal piecewise-constant models of $\alpha(t)$ and $\beta(t)$.  We
use it here to detect variability in the source and the background on
time-scales longer than two weeks.  With the resulting estimates of
$\alpha(t)$ and $\beta(t)$, we can re-scale the photon weights to
effectively remove the slow variability.  We demonstrate this more
explicitly below.

\subsection{Fast Variability With Matched Filters}
\label{sec:matched}

The second method operates in the frequency domain, viz. when $\alpha$
and $\beta$ are periodic with frequency $f$.  The improvement in
log likelihood compared to a steady signal, $\dlogl$, is an exposure
corrected, ``Leahy normalized'' power spectrum, $P(f)\equiv\dlogl(f)$,
whose entries follow a $\chi^2_2$ distribution in the absence of
variability.  K19 demonstrated these properties and showed
that $P(f)$ can be efficiently computed with fast Fourier
transforms.

K19 made the necessary assumption that each Fourier mode is
independent, i.e. the non-diagonal elements of the covariance matrix
in the optimization were ignored.  In general, this is a good
approximation, as off-diagonal elements are typically a few percent of
diagonal elements.  However, it means that the estimator for $P(f)$
inherits some of the same problems as e.g. a Fourier-transform method:
it is a convolution of the true power spectrum with the window
function.

This is a particularly noticeable feature for LAT data because
exposure of the \textit{Fermi} observatory toward a source is a
complicated function of time modulated on the orbital\footnote{Because
of the north/south rocking strategy, the LAT may only view a source
every other orbit, making $\forb/2$ a fundamental modulation
frequency.} ($\sim$96\,min) and precessional ($\sim$53\,d) periods, as
well as by changes in the instrument observing
strategy\footnote{https://fermi.gsfc.nasa.gov/ssc/observations/timeline/}.  An example is shown in Figure \ref{fig:window}.  In the
presence of a signal, the true power spectrum is convolved with this
window function, and power will ``leak'' , appearing in the observed
$P(f)$ at a broad range of frequencies.

From the standpoint of detection, the window function does not affect
the null case (zero flux modulation): the estimators already account for exposure variation,
and there is no signal power to leak, so $P(f)$ follows the $\chi^2$
distribution.  However, leakage can produce apparently periodic
signals from low-frequency noise; and it degrades the sensitivity to
true periodic signals by dispersing the power over the band.  Thus, we
develop here a new method for aggregating the leaked power into a
single estimator, recovering some of the ideal sensitivity that would
pertain with a uniform window function.

% godot implementation is in scaler/CLASS.get_ideal_ps

First, we need to know what $P(f)$ for some true variability
$\alpha(t)$ would look like after it is modified by the complicated
exposure/window function.  For each of the sources, we do this by
computing $s(t)$ using a steady model ($\alpha_0(t)=1$), then modify
this time series by $\alpha(t)$, the desired variability, yielding a
modified prediction $s'(t)$.  Then, we simply evaluate $\dlogl(f)$
using $s'(t)$ in place of the data, $w(t)$.  We call the resulting
power spectrum, $R(f)$, the \textit{response} to distinguish it from
$P(f)$, the power spectrum calculated from the data.  $R(f)$ is the
mean $P(f)$ that would result in the presence of flux variation
$\alpha(t)$.

We show an example $R(f)$ in Figure \ref{fig:harm_example} for a
simulated $f=2$\dinv{} sinusoidal modulation of the emission from
PSR~J0633$+$1746 (Geminga), the second-brightest $\gamma$-ray pulsar.
The dominant signal is at this fundamental frequency, but there are
$<$2\% contributions over a forest of frequencies, and the collective
leaked power substantially exceeds that in the main signal lobe.

This response function reveals where the power leaks and
consequently allows us to capture it.  Specifically, we define the new
statistic 
\begin{equation}
  C^2_D\equiv \sum_i \mathcal{R}(f_i)\,P(f_i) / \sum_i
  \mathcal{R}(f_i),
\end{equation}
which is a weighted sum of the $\delta\log\mathcal{L}$ power spectrum of the data.
Because in the null hypothesis the entries of $P(f_i)$ independently follow a $\chi^2$
distribution, $C^2_D$ follows a $\chi^2_D$ distribution, with
\begin{equation}
  D=\sum_i{\mathcal{R}(f_i)^2}/(\sum_i{\mathcal{R}(f_i)})^2.
\end{equation}
$D$ is a measure of the spectral leakage and can be thought of as the
number of frequency bins a periodic signal is spread over; $D=1$
indicates a perfectly uniform window function\footnote{For a
uniform window function, power leaks as $1/f^2$, but most of the power
is in a single bin.}, while the typical LAT
window function has a $D$$\sim$45, indicating very substantial
spectral leakage.

\begin{figure}
\centering
  \includegraphics[angle=0,width=0.98\linewidth]{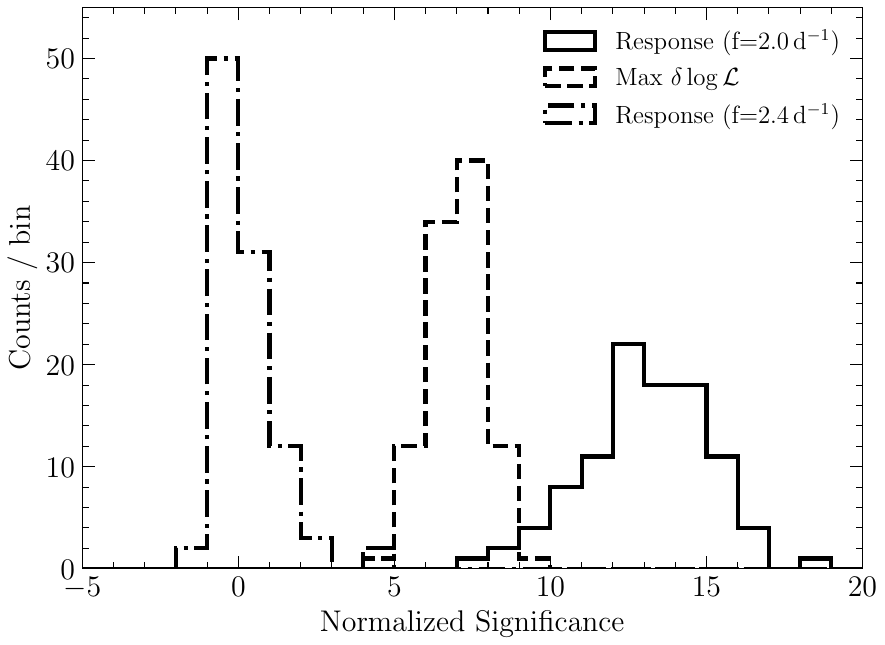}
\caption{\label{fig:win_superiority}The equivalent significances in
  $\sigma$ units for 100 realizations of a simulated period signal
  using the different detection statistics as described in the text.
  The detection statistic using the computed response function for the
  correct frequency results in roughly twice the significance.}
\end{figure}

\begin{figure*}
\centering
  \includegraphics[angle=0,width=0.98\linewidth]{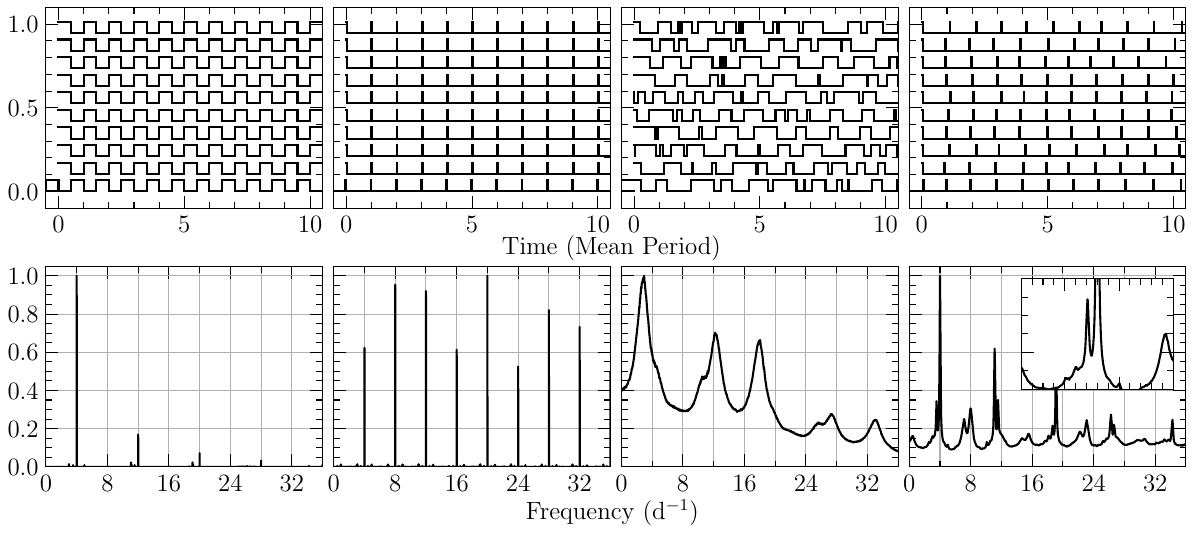}
  \caption{\label{fig:time_freq_example}Four examples of the
  variability process described in \S\ref{sec:model}.  The top four
  panels show 100 cycles of the process, with each 10 cycles wrapped
  to the next higher line.  The bottom four panels show the frequency
  response $R(f)$ for the process, obtained by simulating it
  many times and averaging the resulting Fourier transform.  The
  parameters for the four models are, from left to right: (1)
  $\mathcal{A}=0$,
  $Q\rightarrow\infty$, a uniform square wave, with most of the power
  in the fundamental frequency of $f=4$\dinv{}; (2)
  $|\mathcal{A}|=0.94$,
  $Q\rightarrow\infty$, a very asymmetric but periodic square wave,
  with many strong harmonics; (3) $\mathcal{A}=0$, $Q=1$, a symmetric
  but maximally random process, with possible skipped states, with
  most of the power representing leakage to $f\pm f_{orb}$; (4)
  $\mathcal{A}=0.94$,
  $Q=8$, an asymmetric random process, showing both power concentrated
  in the $f=4$\dinv{} fundamental and harmonics, and leakage.  In the
  later two cases, the randomness of the process smears out the
  periodic structure.  In (4), the increased smearing
  of high-frequency features is clear.  The inset
  shows the primary peak ($0.5\,d^{-1} < f < 7.5\,d^{-1}$) that comprises
  a mixture of harmonics of different widths and amplitudes.}
\end{figure*}

% godot implementation is in simulate/sim_harm
If the trial function used to compute $R(f)$ is close to the true
signal, the matched filter substantially improves the detection
significance.
To demonstrate this, we simulate 100 realizations of $f=2$\dinv{}
sinusoidal modulations in the flux of Geminga with 1\% amplitude.
For each simulation, we compute the significance using (1) the
$\chi^2_2$ calibration for the peak of the raw PSD; (2) $C^2_D$ using
the response for a $f=2.0$\dinv{} sinusoid; and (3) $C^2_D$ for a
$f=2.4$\dinv{} signal.  We show the resulting significance
distributions in Figure \ref{fig:win_superiority}.  By this measure,
using the correct $R(f)$ provides roughly twice the significance!
Conversely, when using the incorrect $f=2.4$\dinv{} response, the
significances are consistent with 0 because this $R(f)$ heavily
weights the parts of $P(f)$ without much power from the simulated
signal.

This, then, comprises the fast variability search strategy: construct
$R(f)$ for various flavors of the stochastic process outlined in
\S\ref{sec:model} and evaluate evidence for it using $C^2_D$.  We show
some example realizations in the time domain along with $R(f)$ in
Figure \ref{fig:time_freq_example}.  To compute these response
functions, we simulated the process many times and averaged the
resulting power spectra $P(f)$.  The left panels show periodic
processes, while the right panels show examples of quasi-periodicity.

%Note to myself: for a given Tf,Tb,Wf,Wb, the width of the features in
%the window function depends on the exact definition of Q.  So e.g.
%suppose Tf=Tb=11800.  Wf=360,Wb=0 yields a much broader feature set
%than Wf=Wb=180.  This can be understand pretty simply by the variance
%of the uniform distribution: the variance in the first case is
%1/12*360**2, in the second (1/12*180**2 + 1/12*180**2) =
%1/12*360**2/2.

\section{Data Preparation}
\label{sec:data}

We use the pulsar timing methodlogy described in \citet{Ajello22} to
process raw LAT data.  In brief, we selected P8R3
\citep{Atwood13,Bruel18} Source-class photons within 3\arcdeg\, of
each pulsar, apply standard quality filtering, and compute weights
with the instrument response appropriate for each PSF event type and
the 4FGL sky model.  The data spans about 15.6\,yr, from MJD 54682 to
60352.  This is the ingredient for the weights time series, $w(t,E)$.

We compute the exposure term $s(t,E)$ using the \texttt{godot}
package\footnote{https://github.com/kerrm/godot}, which was originally
developed to support K19.  We have implemented several
improvements for this project, including: a correction for the trigger
efficiency, based on the livetime; a switch to evaluation of the
exposure at the centers of time bins\footnote{The FT2 file
tabulates the position and attitude of the Fermi spacecraft at the
beginning of 30\,s intervals but provides the livetime integrated over
the interval. Interpolating the attitude to the bin center provides a
more precise evaluation of the instantaneous exposure, an improvement
that is obvious for bright sources.}; support for PSF event types;
and correction of the source counts for the finite aperture size.
These features were primarily and iteratively introduced to reduce
spurious signals from inaccurate exposure calculations appearing in
the $P(f)$ for bright pulsars.  We analyze the data/model agreement in
Appendix \ref{sec:exposure}.

All calculations are carried out in the frame of the solar system
barycenter.  We do not account for the motion of the pulsar in binary
systems, which would degrade the coherence for variability with
timescales comparable to and longer than the binary period (typically
at least 1 day).

\begin{figure*}
\centering
  \includegraphics[angle=0,width=0.98\linewidth]{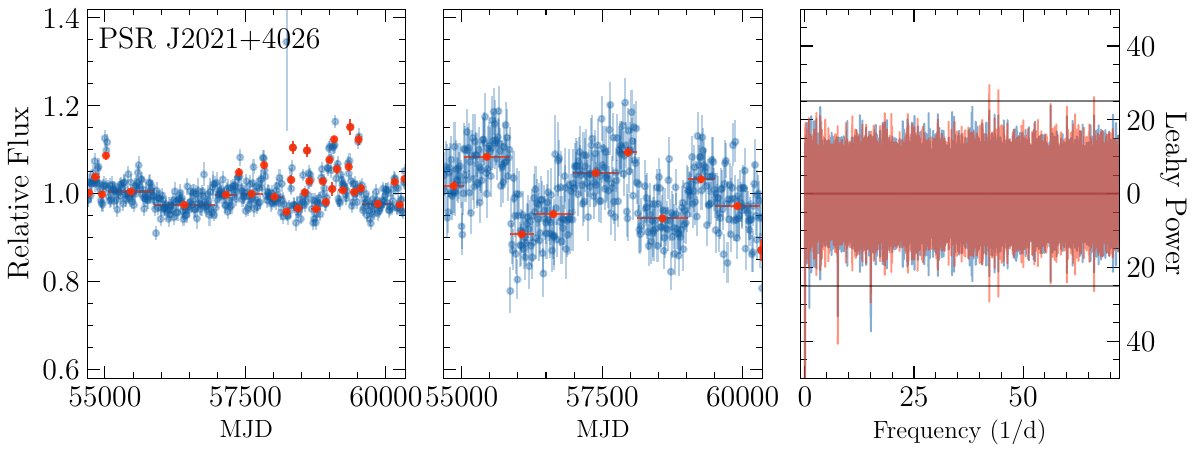}
  \caption{\label{fig:j2021_preprocess}An example of using the
  BB algorithm to identify---and filter---long-term
  variability.  Left: variations in the total
  background driven by activity of Cygnux X-3.
  Center: variations in the source intensity after
  correction for background variations; these are the state changes of
  PSR~J2021$+$4026.  Right: the fixed-background power spectrum $P_0(f)$ in blue and profiled-background $P_1(f)$ in red.  The downward
  (upward) spectrum is before (after) the correction of the photon
  weights for the long-term variability.}
\end{figure*}

\subsection{Sample Selection}

We select the known $\gamma$-ray pulsars those that have a 4FGL-DR4
\citep{4FGL_DR4} ``Test Statistic'' surpassing 1,000.  This yields of
sample of 115 bright pulsars, comprising 70 unrecycled pulsars and 45
millisecond pulsars (MSPs).

\section{Slow Variability}
\label{sec:slow}

The low-frequency/slow variability analysis comprises the following
steps.  We use BB to make an initial estimate of $b(t)$, using
two-week intervals and an exponential prior on the number of change
points $\log n_{cp}=-10$.    This relatively conservative prior
prioritizes a simple model.  We use the estimator for $b(t)$ to
re-weight the photons, $w\rightarrow w/\left[w+b(t)(1-w)\right]$,
which filters this variability, and then estimate $s(t)$ with the same
BB procedure.  The piecewise estimator of $s(t)$ is essentially the
variability test.

Before discussing results for $s(t)$, we describe the ``handoff'' of
the data to the fast variability analysis.  The two-week cadence sets the boundary between the slow and fast variability analyses, as faster variability will be unresolved by BB, whereas slower variability will be filtered out by the handoff process.  First, we perform a second
re-weighting of the photons with $s(t)$ to remove slow source
variability: $w\rightarrow s(t)w/\left[s(t)w + (1-w)\right]$.
This removes the bulk of the signal that could leak to higher
frequencies (see Figure \ref{fig:window}).
We further filter known periodic signals.  Intrinsic
variability timescales include 1\,yr, the S/C precessional period,
$\approx$53.1\dinv{}, and the sidereal day, $\approx$ 1.0028\dinv{}.
For sources neighboring the $\gamma$-ray binaries 1FGL\,J1018.6$-$5856
\citep{Ackermann12_j1018},
LS\,5039, LS I +61 303, and Cygnus X-3, we filter four
harmonics of the orbital periods of those systems.  In all cases, to
avoid spectral leakage, this filter is applied in the time domain by
directly maximizing a linearized likelihood function to estimate the
signal amplitude and phase.  

We demonstrate this detection and filtering approach in Figure
\ref{fig:j2021_preprocess}, which shows the results for
PSR~J2021$+$4026.  The rightmost panel shows that before filtering,
the slow variability in both the background and the source contribute
substantial power at the spacecraft orbital period, i.e. there is spectral
leakage.  By constructing a time-domain model with BB and re-weighting
the photons, the spectral leakage is greatly reduced, resulting in
clean power spectra.

\subsection{Individual Results}

While in principle the presence of even one flux change point indicates
statistically significant variability, in practice it is not always
possible to exactly separate the variations of the pulsar and the
background.  We thus visually inspected the results for every pulsar,
generally finding no strong evidence for new slow variability.
We detail a few of the cases here.

\subsubsection{J0205$+$6449}
The light curve of PSR~J0205$+$6449, an energetic pulsar associated
with the 3C\,58 plerion, has a statistically significant flare
beginning around MJD 57827 and lasting for a few weeks.  Analysis of
LAT data restricted to the off-pulse of the pulsar indicates that the
cause is a likely flare of the blazar J0209$+$6437, about 0.5\arcdeg
away (priv. comm. from C.~C.~Cheung and T.~J.~Johnson.)

\subsubsection{PSR\,J0534$+$2200}

The Crab pulsar wind nebula is co-located with the Crab pulsar and its
emission dominates the high-energy spectrum above about 10\,GeV.  The
X-ray amplitude of the nebula burbles \citep{Wilson-Hodge_11}, and it
occasionally flares dramatically in the $\gamma$-ray band with sub-day
variability and a spectrum that cuts off at $\sim$1\,GeV
\citep{Abdo11_crab,Buehler12}.  This spectral similarity makes the
separation of the pulsar and PWN components by photon weights poor.
Because the PWN varies at low levels constantly, there is no
quiescence period to select.  Consequently, even though this is one of
the brightest pulsars in the sample, we exclude it from further
consideration.

\subsubsection{PSR\,J1227-4853}

This source is a transitional redback MSP \citep{Roy15} that is a
bright $\gamma$-ray pulsar in the rotational powered state
\citep{Johnson15}.  Analysis of the light curve reveals two steps
during the accretion phase, and a steady state since MJD 56293 during
the rotation-powered state.  We exclude the interval before this point
from the fast variability analysis.

\subsubsection{PSR\,J2032$+$4127}

This pulsar is only 0.5\degr{} from Cygnus X-3, which occasionally
emits GeV flares modulated on the 4.8-hr period
\citep{Abdo09_cygx3,Kerr19}.  This separation is comparable to the LAT
PSF width below a few GeV, so these sources are confused.  The 4.8-hr
signal is explicitly removed by the filtering procedure, but the
brightest Cygnus X-3 flares around MJD 59000 produces apparent
low-level variability in the BB light curve.

\subsection{Single Steps}
Eight of the pulsars display a single step at the beginning or end
of the data span, with a typical length of 10--20\% of the data.
These cases do not appear to be affected by background contamination,
so they could indicate real state changes.  However, these single
steps are more likely to arise by chance compared to a fully-contained
state because the penalty of the prior on the number of change points
is only incurred once.  Further, if most of these were real, we would
expect to see at least a few enclosed state changes.  Of these cases,
the ones that seem most likely to be genuine occur in pulsars
J0744$+$2525, J1057$-$5226, and J1957$+$5033.
\begin{figure}
\centering
  \includegraphics[angle=0,width=0.98\linewidth]{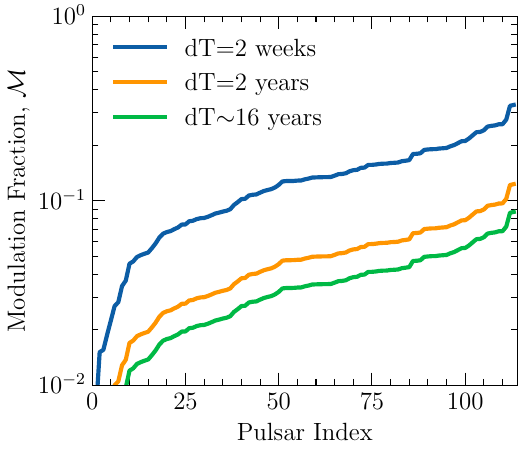}
  \caption{\label{fig:slow_sens}The 2$\sigma$ confidence upper limit
  on modulation fraction (Eq. \ref{eq:M}) estimated per
  \S\ref{sec:slow_sens}.  The estimates are shown for three
  representative timescales for each pulsar after sorting in order of
  decreasing source brightness.  Modulation fractions $>$3--10\% are ruled out for most pulsars at the longer timescales.}
\end{figure}

\subsection{Sensitivity Estimate}
\label{sec:slow_sens}
We can estimate the variability level required for single-trial detection as
follows.  The weighted photon rate $w(t)$ from each pulsar determines
the precision, $\sigma_F$, with which the flux can be measured within each
$t_{cell}=14$\,d cell of the BB light curve.  We estimate $\sigma_F$ as the mean observed uncertainty.  Suppose
there is an additional
variability that increases the observed flux scatter that has a standard deviation $\sigma_{v}$ and operates on a characteristic
timescale $t_{v}\geq t_{cell}$.  With a total data length of $T$,
there are on average $N_{v}=T/t_{v}$ states, each comprising on
average $n_{v}=t_{v}/t_{cell}$ cells.  If we were to evaluate
$\delta\chi^2$ from the difference in the null hypothesis,
$\alpha(t)=1$, and the alternative hypothesis provided by $\alpha(t)$
from the best-fit BB light curve, then in the null hypothesis
$\left\langle\delta\chi^2\right\rangle=\sum_{i=1}^{N_{v}}
\sigma_{v}^2/(\sigma_F^2/n_{v})
= N_{v} n_{v}\sigma_{v}^2/\sigma_F^2$.  The average variance is
$2N_{v}$, so in $\sigma$ units, this is $n_{v}\sqrt{N_{v}/2}
\sigma_{v}^2/\sigma_F^2$.  Finally, setting this to a detection
threshold $\kappa$, and reverting to timespans, we have the typical
variability level required to surprass a
$\kappa\sigma$ sensitivity threshold: $\sigma_{v}\geq
\sigma_F\left(2\kappa^2
t_{cell}^2\,T^{-1}\,t_v^{-1})\right)^{\frac{1}{4}}$.  Since this is a fractional modulation, for ease of comparison, we label it with the same symbol $\mathcal{M}$ we use for the two-state process.  Adopting $\kappa=2$, we report the limits in Figure \ref{fig:slow_sens}.

This approach breaks down when $t_{v}$ approaches $T$.  We thus also
estimate a limit for a very slow process that has at most one change
point.  Taking this to occur 25\% of the way into the data span, the
sensitivity is simply the precision of this 4-year block,
$\sigma_{var}\geq{t_{cell}\,t_{v}}^{-\frac{1}{2}}\sigma_{F}\approx0.14\,\sigma_F$,
which is also shown in Figure \ref{fig:slow_sens}.

\section{Fast Variability}
\label{sec:fast}

We now present a search for fast variability using the matched filter
technique outlined in \S\ref{sec:matched} using power spectra $P(f)$
produced from time series of photon weights that are filtered for
slow variability as described in \S\ref{sec:slow}.

First, we note that obtaining the exact matched filter $R(f)$ requires
simulating hundreds of realizations of the process, which is
prohibitively expensive for a search over the parameter space of
asymmetry $\mathcal{A}$, randomness $Q$, and time-scale/frequency $f$. 
Instead, we have devised a method for constructing approximations of
$R(f)$ that is presented in Appendix \ref{sec:approximate_filters}.
To recapitulate, we choose four representative values with ratios
$T_f/T_b$ of 1, 3, 10, and 30, corresponding to 
$\mathcal{A}=0$, 0.5, 0.82, and 0.94.  Because the shape of $R(f)$ is
invariant under $T_f/T_b\rightarrow T_b/T_f$, these values also
include $\mathcal{A}=0$, $-0.5$, $-0.82$, and $-0.94$.  We then use
the relation between the width of spectral features and the
quality parameter, $\sigma_f\propto
f/Q^2$, to overlay gaussians of the correct amplitude and width at the
harmonics of the square wave with frequency $f$.  This effectively
re-maps the search space to one in $\mathcal{A}$, $\sigma_f$, and $f$.

More specifically, the search is carried out as follows.  We bin the
data into 300\,s cells and compute the various power spectra $P(f)$ up
to a limiting frequency 72\,d$^{-1}$, corresponding to a sampling time
of 600\,s or 10 minutes.  (Because terms in $2f$ are needed
to calibrate $P(f)$, only the lower half of the band is
available; see K19.)  For each of
the four representative $\mathcal{A}$ windows, we search widths
$\sigma_f=2^i$ with $i$ ranging from 0--15, i.e. from $W=1$ to 32,768
spectral bins.
This maximum width is about 10\% of the total bandwidth.
For each value of $W$, we form an approximate template at $f=0$.
Because templates with $f=0$ are related by translation and width
scaling, we carry out the frequency search by simply shift the template to higher
frequencies and
evaluating $C^2_D$ for each step.  (See
Appendix \ref{sec:approximate_filters} for consideration of boundary
conditions.)  The frequency step size is
$W/2$.  The final results of the search are the values $C^2_D$ and
$D$, which can be converted to chance probabilities.  These statistics
naturally reflect the inhomogeneous sensitivity of the search.  (E.g.,
bursty processes with short time-scales are more poorly constrained because
fewer harmonics fit within the band.)

In order to identify contamination from background variations, we
carry out parallel searches using two realizations of $P(f)$ as input:
$P_0(f)$ assumes no periodic variations of the background, while
$P_1(f)$ is a profile likelihood spectrum where the background
variations are fixed to their maximum likelihood values.  $P_0(f)$ is
more sensitive to both source and background variations, while
$P_1(f)$ trades some sensitivity to mitigate background contamination.
See K19 for more details.

\begin{figure*}
\centering
  \includegraphics[angle=0,width=0.98\linewidth]{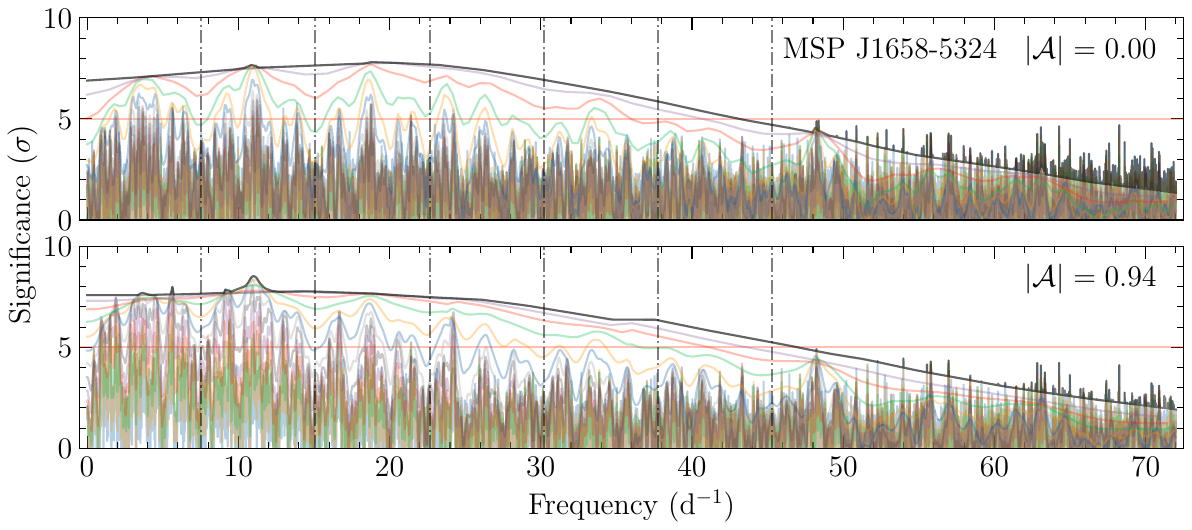}
  \caption{\label{fig:j1658}The significance of fast variability for
  PSR~J1658$-$5324 over the targeted frequency range.  Each line
  indicates the results for a matched filter of increasing width,
  i.e. an increasing level of randomness in the duration of each
  state.  The maximum signal in any filter is shown by the black line,
  which for most frequencies is the widest filter, corresponding to a
  width of about 10\,d$^{-1}$.}
\end{figure*}

We construct a spectrum of the detection statistic by selecting the
maximum value of  $C^2_D$ obtained over all values of
$\mathcal{A}$ and $W$ for each frequency bin.  We then take the
bin-by-bin minimum of the two spectra produced from $P_0$ and $P_1$.
An example of this spectrum is shown in Figure \ref{fig:j1658}.
Finally, we take the maximum value of the spectrum as the raw
detection statistic for the search.

The $C^2_D$ values are not statistically independent over $f$,
$\mathcal{A}$, and the 15 hierarchical values of $W$.  However, we can
conservatively estimate the total number of trials as $7\times10^8$,
being the product of $n_{freq} = 4\times10^5$, $n_{pulsar}=115$,
$n_{\mathcal{A}}=4$, and $\log_2{n_W}=\log_2{15}$.  Requiring a 1\%
chance of a false positive detection suggests a threshold of
6.7$\sigma$.  Indeed, the majority of maximum detection statistic
values lie in the range 5--6.5$\sigma$.  To account for minor
contributions of leaked background power, we adopt a 7$\sigma$
threshold for variability candidates.  We discuss the sources that pass
this threshold below, along with two notable nondetections.  Except
where otherwise noted, we see no link between the frequencies with
peak significance and any characteristic instrumental frequency,
particularly $\forb{}$.

\subsection{Individual Results}

\subsubsection{J0218$+$4232}
The test statistic for this pulsar peaks at 7.6\,$\sigma$ at
frequencies of 39\,\dinv{}, which does not correspond to an obvious
leakage frequency.  On the other hand, the bright and variable 3C~66A
blazar is about 1\arcdeg{} away, which suggests a possible systematic
origin of the apparent variability.

\subsubsection{PSR\,J0613$-$0200}
J0613-0200 has a 7.3$\sigma$ excess at about 33\,\dinv{}.  Because
there is no obvious background contamination, this could correspond to
a bona fide quasi-periodic variability, warranting future (perhaps
spin-phase resolved) analysis.

\subsubsection{PSR\,J0835$-$4510}
Although there is no variability signal in excess of 7$\sigma$, we
mention this source explicitly because Vela is so bright that its
nondetection gives confidence that the fast variability search is not
limited by systematic deficiencies in the exposure calculation.

\subsubsection{PSR\,J1658$-$5324}
This pulsar, which was found in targeted radio searches of unidentified
LAT sources \citep{Kerr12} has a variability significance of
8.5$\sigma$ obtained with a very wide matched filter--essentially the
entire band---indicating it would originate in process without a well
defined timescale.  The spectrum for this pulsar is shown as an example in Figure \ref{fig:j1658}; spectra for the remaining pulsars discussed here are shown in Appendix \ref{sec:app_plots}.

\subsubsection{PSR\,J1816$+$4510}

We excise data in the MJD intervals 57350--57800 and 58500--58700 due
to an exceptionally bright flaring background source.  There is no
residual slow variability apparent in the source, but the fast
variability analysis yields a 7.3$\sigma$ excess peaked at 16\dinv{},
which is likely due to background contamination.

\subsubsection{PSR\,J2021$+$4026}
As the only known (flux) state-changing $\gamma$-ray pulsar, it is
notable that this pulsar exhibits no detectable fast variability.
Moreover, the nondetection further demonstrates the effectiveness of the
two-prong approach in separating slow variability (from the pulsar and
from the background source Cygnus X-3) and fast variability.

\subsubsection{PSR\,J2032$+$4127}
There is an excess at 15\,\dinv{}$\approx\forb$, consistent with leakage of low-frequency
contamination from Cygnus X-3, indicating the limitations of removing
background variations in extreme cases.

\subsubsection{PSR~J2256$-$1024}
There is a narrow peak at about 34\,\dinv{}.  There is also a very
strong variable background source active from MJD 57500 to
59,300.  This is likely the source of the signal, though it is
difficult to understand how it could produce such a narrow feature at
that particular frequency.  This is a source that would benefit from
improved weights computation, viz. one with a more detailed
variability model for the background source.

\subsection{Eclipses}

Some of the binary MSPs in the sample were noted by \citet{Clark23b} to
exhibit $\gamma$-ray eclipses.  Our model accommodates such signals,
but we do not recover them because the matched filter methods used can
only partially recover leaked signals.  The time-domain methods of
\citet{Clark23b} avoid spectral leakage and are also able to target the
phase of superior conjunction, when eclipses are expected.

\subsection{Sensitivity Estimation}
To estimate the sensitivity to fast variability, we construct the exact $R(f)$ for a range
of $\mathcal{A}$ and $\sigma_f$/$W$ parameters with the maximal
$\mathcal{M}=1$.  From Eq. \ref{eq:total_power}, the total time-domain
power is $\sigma_2=(1-\mathcal{A})/(1+\mathcal{A})$.  Recall that
$R(f)$ is the expectation value of the log likelihood spectrum,
$P(f)$.  Consequently, letting $P(f)\rightarrow R(f)$ yields the
expectation value of $C_{D}(\mathcal{M}=1)$. Conversely, the
no-modulation expectation value is $C_{D}(\mathcal{M}=0)=D$.  Finally,
for any given $D$, we can use the $\chi^2$ distribution to determine
the detection threshold for $C_{D}(N\sigma)$ with a tail probability
corresponding to $N\sigma$.  Since $C_D$ is linear in $\sigma^2$, the
total power required for an $N\sigma$ detection is 
\begin{equation}
  \sigma^2_{N\sigma} =
  \frac{1-\mathcal{A}}{1+\mathcal{A}}\frac{C_{D}(N\sigma)-D}{C_{D}(\mathcal{M}=1)-D}.
\end{equation}
Finally, we solve Eq. \ref{eq:total_power} for $\mathcal{M}(N\sigma)$, the
fractional modulation required for $N\sigma$ detection of the process
with parameters $\mathcal{A}$ and $W$.

\begin{figure*}
\centering
  \includegraphics[angle=0,width=0.98\linewidth]{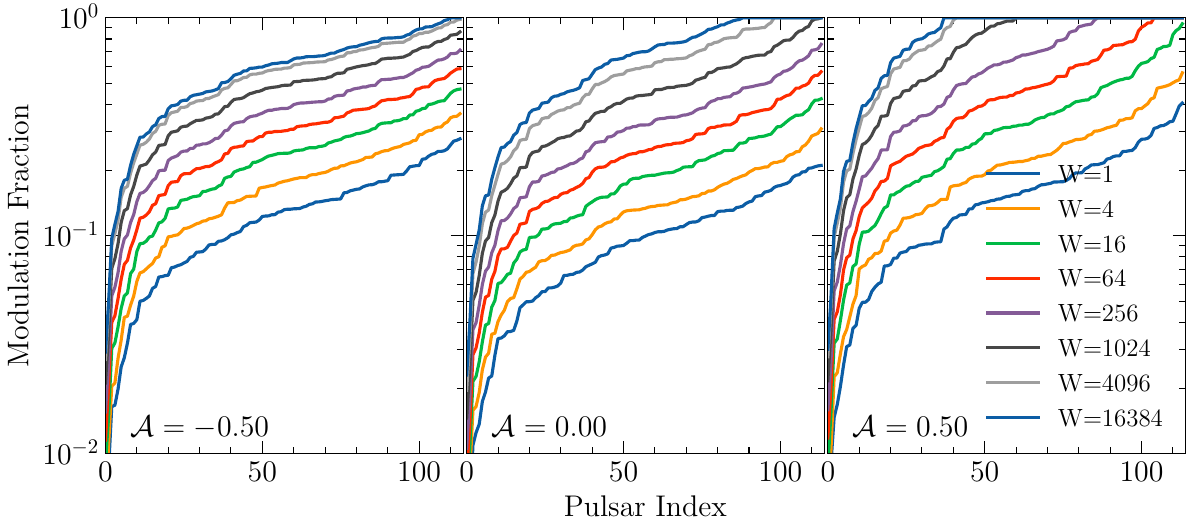}
  \caption{\label{fig:fast_sens}The modulation fraction $\mathcal{M}$ required to exceed a 5$\sigma$
  detection threshold. The curves shows the constraints for the
  indicated value of $W$, which is the width of spectral features ($\sigma_f$) expressed in frequency bins.  $W$ is
  related to the process randomness and timescale.  The left (right) panel
  shows pulsars that spend three times longer in the faint (bright)
  state, while the center panel features equal distribution of
  time.  The $\mathcal{A}=0.5$ constraints are weaker due to the lower total variance.  In general, nearly-periodic processes
  are constrained to $<$10--20\% modulation for the full sample.  For even nearly random
  processes, nulling is ruled out for the full sample for
  the more constrained $\mathcal{A}<=0$ models.  For the
  brightest pulsars, the modulation fraction is restricted to
  $\lesssim$10\% regardless of the degree of quasi-periodicity.
  }  
\end{figure*}

The value of $\mathcal{M}(N\sigma)$ depends slightly on $f$ because
processes with a faster intrinsic timescale have fewer in-band
harmonics.  However, the effect is generally 10\% or less, so for
simplicity we consider sensitivity at the central frequency
$f/f_{max}=0.5$.  The sensitivity for three typical values of
$\mathcal{A}$ for each pulsar in the sample is shown in Figure
\ref{fig:fast_sens}. 

\section{Discussion and Future Work}
\label{sec:discussion}

To summarize, the goal of this work was to search for and characterize
flux variability associated with random state changes with timescales
ranging from about 10 minutes to a few days.  To achieve this, we used
a two-stage analysis that first identified and filtered variable of
the background and pulsar on slower ($\geq$2 weeks) timescales, then
used a matched filtering technique to search for random processes in
the spectral domain.

There are some caveats: first, we have only analyzed the 115 brightest
of about 300 pulsars \citep{Smith23}.  We adopted the threshold of
TS$=$1000 because of our focus on fast modulation: as shown in Figure
\ref{fig:fast_sens}, it would be impossible to difficult to detect
anything other than nearly-periodic modulation in fainter pulsars.
However, such extreme variability could exist in the fainter sample,
as could less extreme, slower variability, motivating future
examination of the fainter sample.

Second, our methodology does require at least an element of
quasiperiodicity in order to manifest a $P(f)$ that is not flat, so
a state switching process with an unbounded maximum state residence
time would evade detection.

With these caveats, however, we have the following result: $\gamma$-ray
pulsar flux state changes are neither widespread nor strong.  Slow
flux variability is limited to the 10\% level over the entire sample.
This means that PSR~J2021$+$4026 is unique.

Constraints are weaker for fast variability, but for several pulsars,
we can rule out fast state switches with flux differences of even a
few per cent, regardless of the degree of randomness in the switching
process.  For nearly-coherent state switches, we can rule out
switching with amplitudes more than 10\% over most of the population.
Full nulling is ruled out over most of the sample on all of the
timescales probed.

If we interpret these flux constraints as also limiting the variations
in $\dot{E}$, then we can also state that $\gamma$-ray emitting
pulsars are almost always in a force-free magnetosphere state, with
variations about this equilibrium limited to 1--10\%.

The limits for the brightest pulsars in the sample rule out the upper
end of the 0.1--10\% $\dot{\nu}$ variations observed by \cite{Lyne10},
even though these variations also occur in high-$\dot{E}$ pulsars
\citep{lower_ubiquity_2025} with $\gamma$-ray emitting outer
magnetospheres.  This result could indicate that the physical
mechanism that drives these variations does not operate at short
timescales.  On the other hand, the scatter in the amplitude of
$\dot{\nu}$ variations is large, so a larger sample of strong limits
on $\gamma$-ray variability is needed to ensure that it is likely to
include pulsars with large $\dot{\nu}$ variations.  Analysis of the
timing properties of the $\gamma$-ray sample might also identify such
pulsars.

We found very modest evidence for some single-step flux changes in
pulsar (very slow variability), and in a few pulars we found excess
high-frequency power indicative of switching on timescales faster than
one hour.  The interpretation of these marginal cases is complicated
by potential background contamination.  Thus, the most promising
approach for future work is to incorporate pulse (spin) phase
information.  Using the average pulse profile to further weight
photons associated with the pulse peaks will improve sensitivity by
$>$3$\times$ for typical pulsars in the sample, expanding the sample in which
$\leq$1\% $\dot{\nu}$ variations could be detected.  Incorporation of
spin phase will also make it possible to search for variations in the
pulse profile shape, rather than just the total intensity, and to look
at spindown/pulse profile variations using methods similar to those
used to analyze radio pulsar populations
\citep[e.g.][]{brook_emission-rotation_2016}.  Finally, because the
emission from most pulsars vanishes for some range of the neutron star
rotation, data from this off-pulse can be used to fully remove the
effects of background variations.

%Other idea for future work: "bake in" the known source variability,
%which would improve the weights computation.

\begin{acknowledgements}
The \textit{Fermi} LAT Collaboration acknowledges generous ongoing
support from a number of agencies and institutes that have supported
both the development and the operation of the LAT as well as
scientific data analysis.  These include the National Aeronautics and
Space Administration and the Department of Energy in the United
States, the Commissariat \`a l'Energie Atomique and the Centre
National de la Recherche Scientifique / Institut National de Physique
Nucl\'eaire et de Physique des Particules in France, the Agenzia
Spaziale Italiana and the Istituto Nazionale di Fisica Nucleare in
Italy, the Ministry of Education, Culture, Sports, Science and
Technology (MEXT), High Energy Accelerator Research Organization (KEK)
and Japan Aerospace Exploration Agency (JAXA) in Japan, and the
K.~A.~Wallenberg Foundation, the Swedish Research Council and the
Swedish National Space Board in Sweden.
 
Additional support for science analysis during the operations phase
is gratefully acknowledged from the Istituto Nazionale di Astrofisica
in Italy and the Centre National d'\'Etudes Spatiales in France. This
work performed in part under DOE Contract DE-AC02-76SF00515.

Work at NRL is supported by NASA, in part by Fermi Guest Investigator
grant NNG24OB06A.

\end{acknowledgements}

\facilities{Fermi}

\bibliographystyle{aasjournalv7}
\bibliography{sr,pmag,pmag_obs,pmag_classic,ns_int}

\appendix

\section{Approximating Matched Filters}
\label{sec:approximate_filters}

\begin{figure}
\centering
  \includegraphics[angle=0,width=0.98\linewidth]{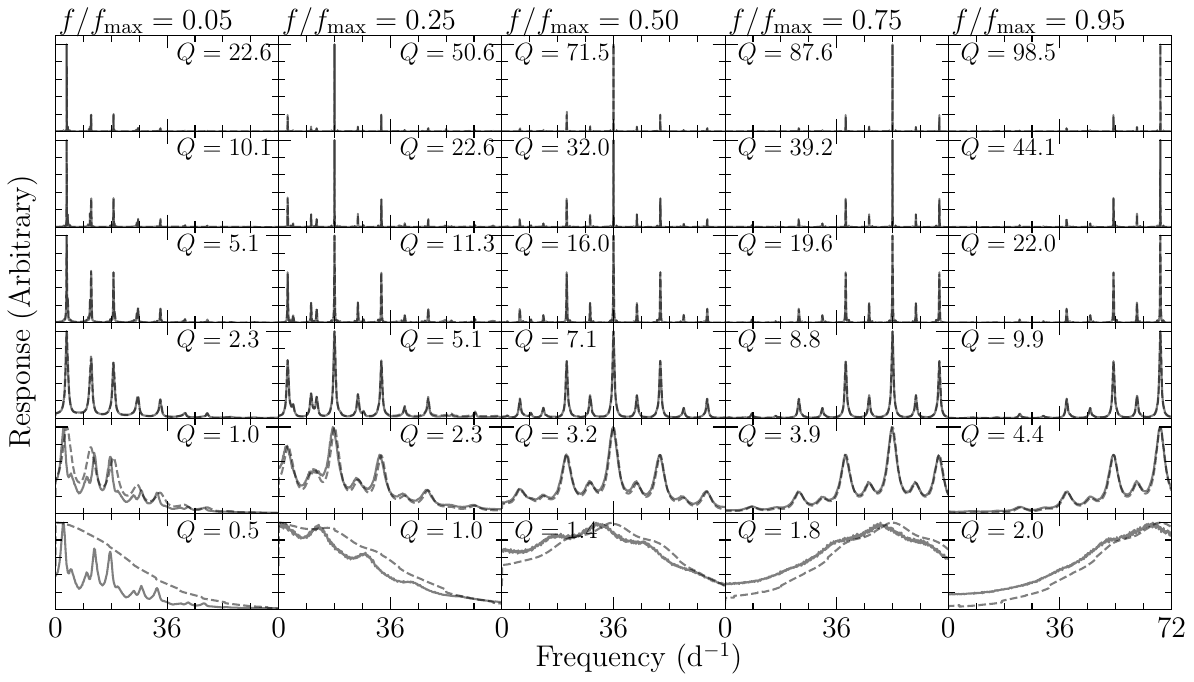}
  \includegraphics[angle=0,width=0.98\linewidth]{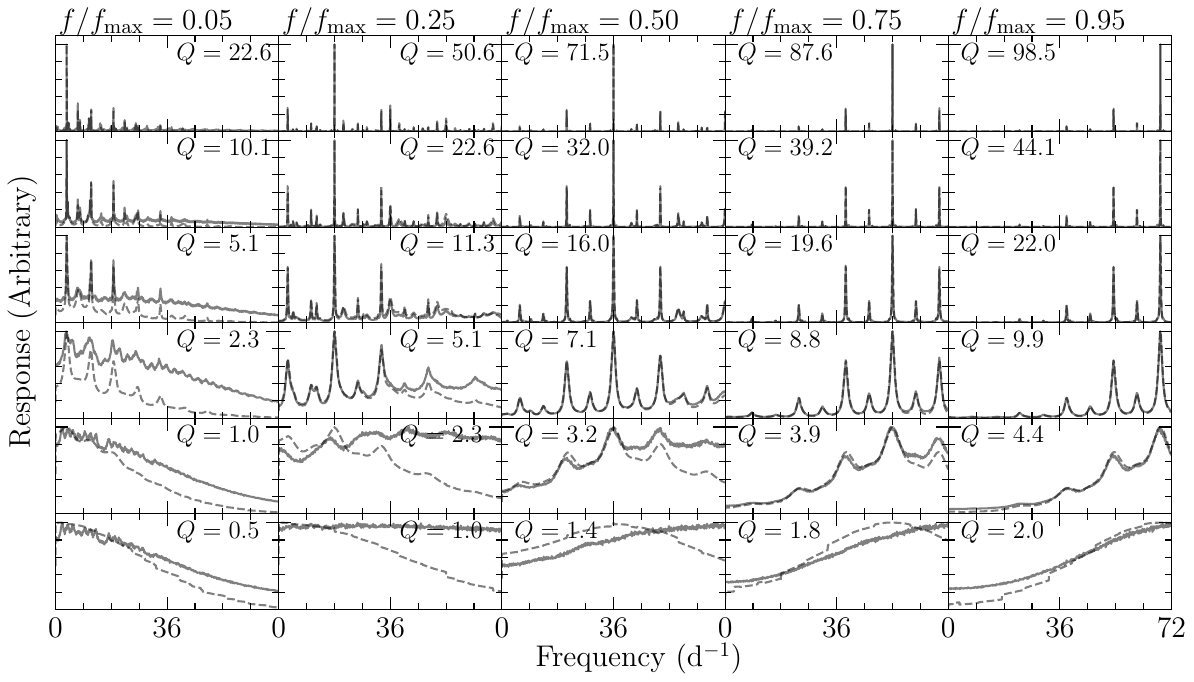}
  \caption{\label{fig:approx_filters}Approximate filters for
  asymmetry $\mathcal{A}=0$ (top) and $|\mathcal{A}|=29/31$ (bottom).  The
  light grey solid line shows the exact $R(f)$, while the dashed line
  shows a close-matching approximate version.  The
  signal frequency increases left to right and is expressed as a
  fraction of the highest frequency of $P(f)$, namely 72\dinv{}.  The
  value of $Q$ is used to estimate the width $W$ of the approximate
  filter, which can be seen to narrow as $f$ increases.  Instances $Q<1$
  are not meaningful but are displayed for completeness.}
\end{figure}

Recall that the degree of randomness in the pulsar variability model
is governed by $Q\equiv(T_f+T_b)/(W_f+W_b)$.  This randomness broadens
the features in the matched filter $R(f)$ as demonstrated in Figure
\ref{fig:time_freq_example}, with a characteristic width $\sigma_f
\propto f/Q^2$.  The relative strengths of the harmonics in the PSD of
the square wave process depend on the asymmetry, $\mathcal{A}$.
(Recall that in a $\mathcal{A}=0$ process, equal time is spent in each
state, while a $\mathcal{A}=1$ ($=-1$) process is always in the bright
(faint) state.)  In general, the shape of the power spectrum only
depends on $|\mathcal{A}|$, but the total power is stronger for
negative values of $\mathcal{A}$.

A low-$\mathcal{A}$ process has most power in the fundamental at $2f$
and the first nonzero harmonic at $4f$.  Thus, an approximate template
can be constructed by superposing two gaussians with widths $W$ and
$2W$ centered on frequencies $2f$ and $4f$.  The only unknown is the
relative amplitudes.

A very bursty $|\mathcal{A}|\rightarrow 1$ process has approximately
uniform power in its harmonics, and the resulting response comprises
narrow and broad spectral features lying atop a broad wedge of pink
noise from harmonics that have been completely smeared out across the
band.  Because the width of the gaussians approximating higher
harmonics increases rapidly, contributions from above the fifth
harmonic or so can be approximated as a pedestal.  As the process
timescale ($1/f$) varies, the absolute widths of the gaussians evolve
according to $\sigma_f \propto f/Q^2$.

Importantly, this relation also enables an efficient search of the
parameter space.  E.g., a template constucted to search at a frequency
$f$ and randomness $Q$ can be shifted to frequency $f'$ by fast
operations like shifts and reflections, where it corresponds to a
process with randomness parameter $Q \sqrt{f'/f}$. Thus, rather than
searching directly over $Q$, we search a grid of $W$ and $f$ that
spans the same range of $Q$ of interest.

We have combined these various ingredients into a recipe for making
reasonable matched filter approximations for four discrete values of
the process asymmetry, $|\mathcal{A}|=0$, $1/2$, $9/11$, and $29/31$.
We tuned the number and amplitude of the gaussians and pedestal
components in each case by comparing them to the exact realizations of
$R(f)$ generated by simulation.  We show the approximate templates for
the two extrema of $|\mathcal{A}|=0$ (square wave) and
$|\mathcal{A}|=29/31$ (impulse train) in Figure
\ref{fig:approx_filters}.  The agreement is generally excellent,
indicating that little search efficiency is lost when using the
approximations.

\section{On Exposure Calculations}
\label{sec:exposure}

\begin{figure*}
\centering
  \includegraphics[angle=0,width=0.98\linewidth]{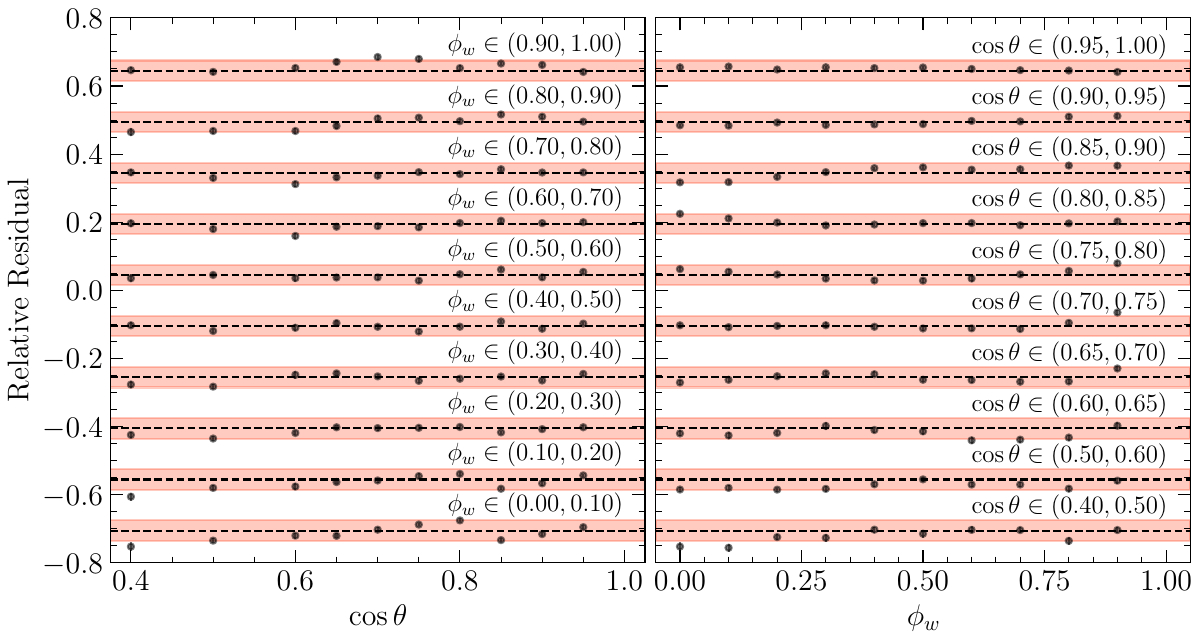}
  \caption{\label{fig:exposure_check}The observed versus computed
  counts for PSR~J0835$-$4510 (Vela) for various selection of the
  incidence angle from boresight, $\theta$ and the wrapped azimuthal angle
  $\phi_w\equiv 2\left| \phi\pmod{\frac{\pi}{2}}\frac{2}{\pi}
  -\frac{1}{2}\right|$, which ranges from 1 (along a face) to 0 (at a
  corner). The shaded bands indicate $\pm$3\%.  Each set of
  residuals is centered on 0 but offset by 0.15 for display.}
\end{figure*}

To demonstrate the accuracy of the exposure calculation through the
full \texttt{godot} toolchain, we have tabulated the difference
between the observed and predicted weights $w$ and $s$ for the
brightest pulsar, PSR~J0835$-$4510 (Vela), as a function of the
instrumental coordinates $\phi$ (azimuth) and $\theta$ (polar angle),
and plotted them in Figure \ref{fig:exposure_check}.
It is a testament to the LAT team that the agreement is generally at
the per cent level!

\section{Additional spectra}
\label{sec:app_plots}

\begin{figure*}
\centering
  \includegraphics[angle=0,width=0.49\linewidth]{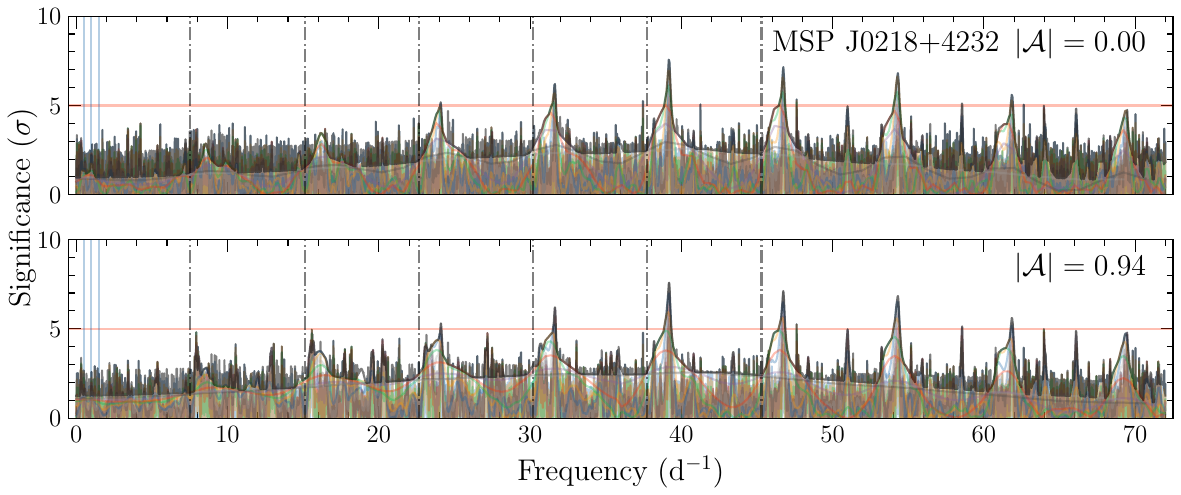}
  \includegraphics[angle=0,width=0.49\linewidth]{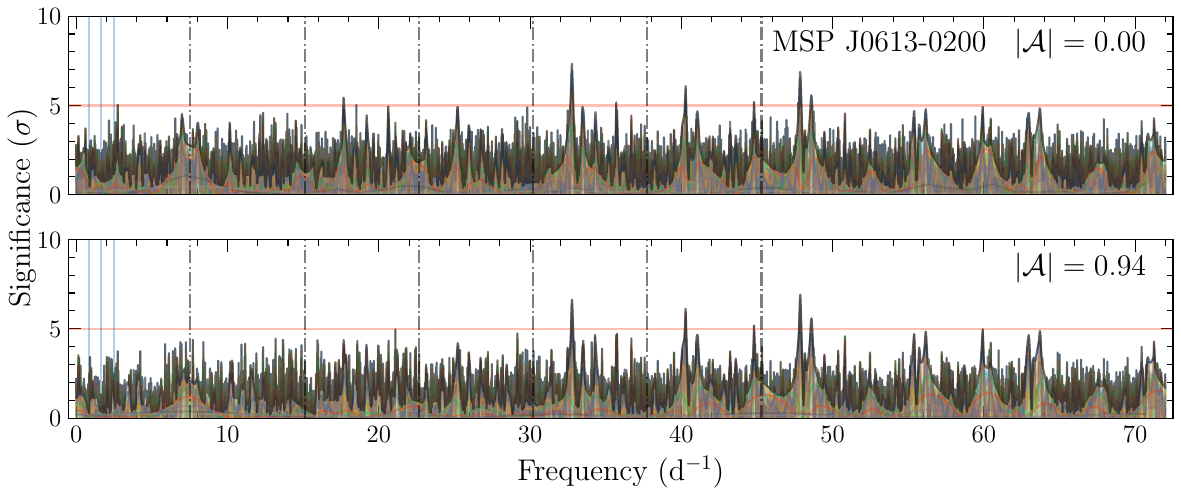}
  \includegraphics[angle=0,width=0.49\linewidth]{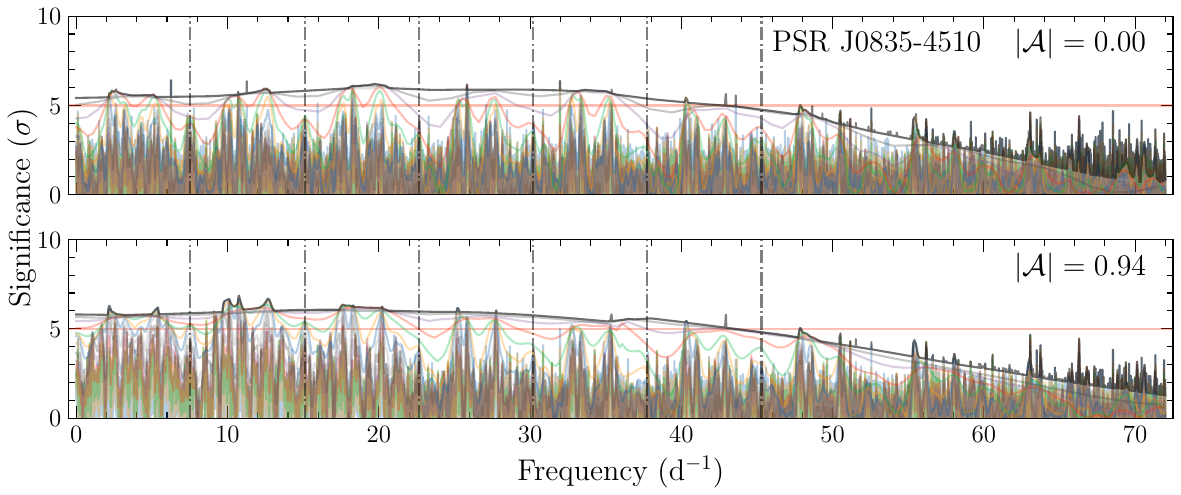}
  \includegraphics[angle=0,width=0.49\linewidth]{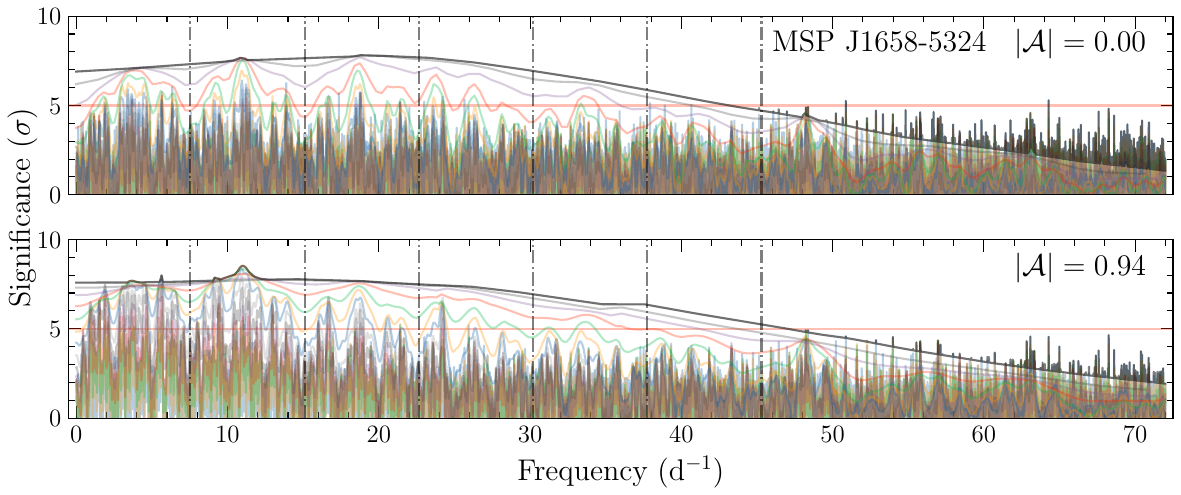}
  \includegraphics[angle=0,width=0.49\linewidth]{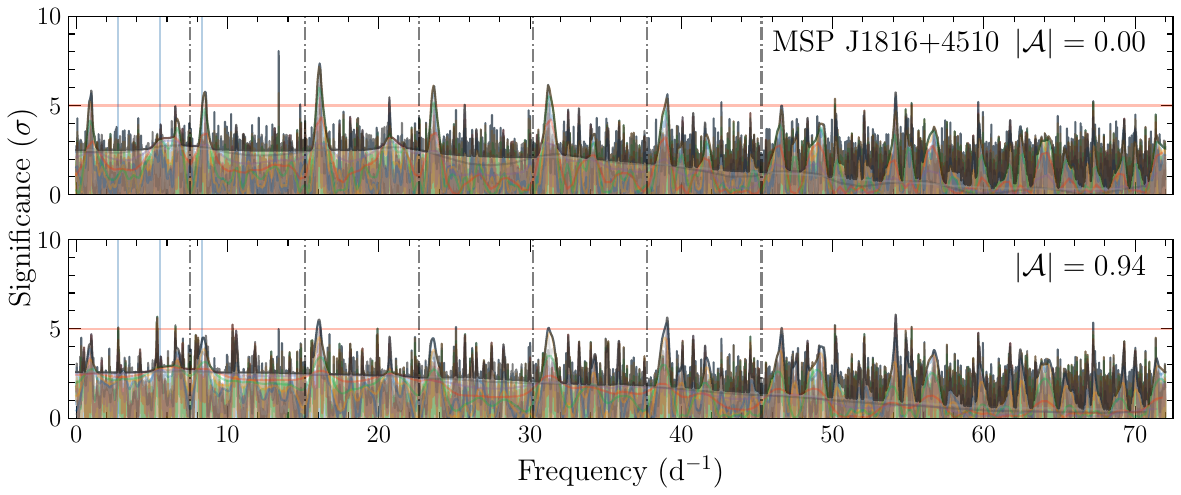}
  \includegraphics[angle=0,width=0.49\linewidth]{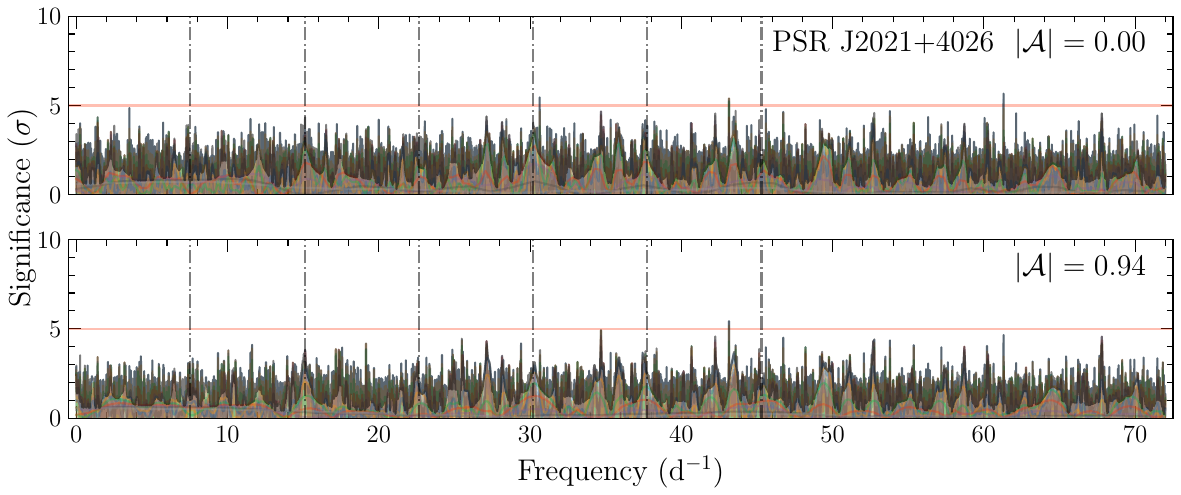}
  \includegraphics[angle=0,width=0.49\linewidth]{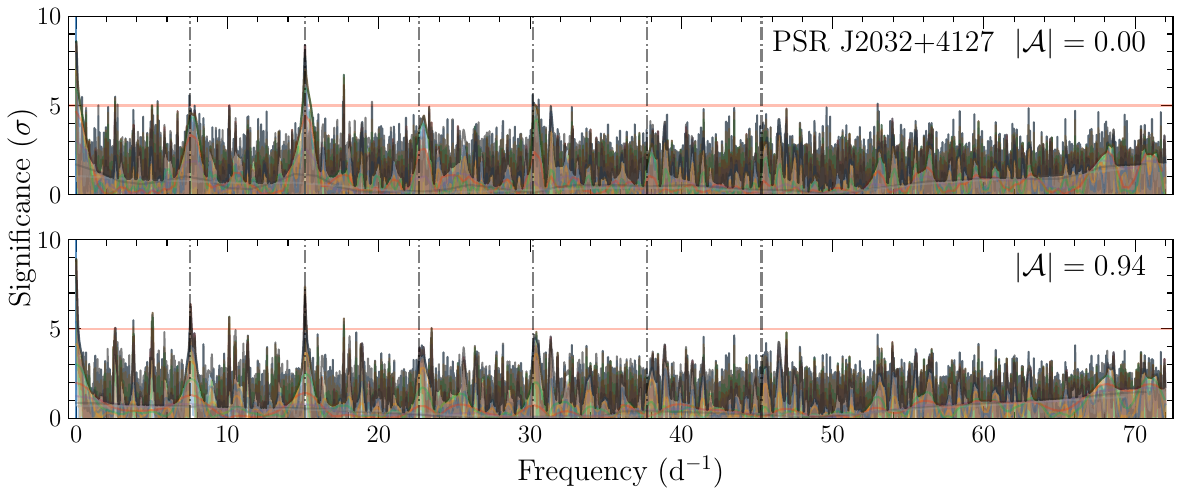}
  \includegraphics[angle=0,width=0.49\linewidth]{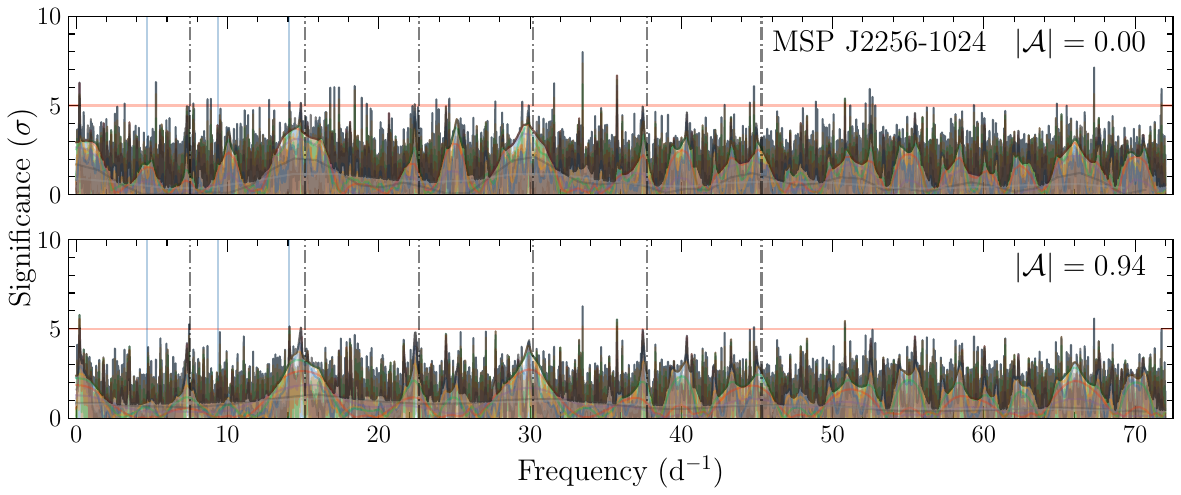}
  \caption{\label{fig:spectra}As Figure \ref{fig:j1658}, for additional pulsars as labeled.}
\end{figure*}

Figure \ref{fig:spectra} presents spectra for all pulsars discussed in \S\ref{sec:fast}, which includes candidates with $>7\sigma$ fast variability as well as noteworthy nondetections.

\end{document}